\documentclass[aps,pra,superscriptaddress,nofootinbib,longbibliography,notitlepage,
11pt,
floatfix]{revtex4-1}
\usepackage{amsmath}
\usepackage{amsthm}
\usepackage{amsfonts}
\usepackage{amssymb}
\usepackage{hyperref}
\usepackage{color}
\usepackage{microtype}

\newcommand{\onenorm}[1]{{\left\|{#1}\right\|_1}}

\newcommand{\cS}{{\mathcal{S}}}
\newcommand{\nbar}{\overline{n_\mathrm{out}}}
\newcommand{\nout}{{n_\mathrm{out}}}
\newcommand{\ebar}{\overline{\epsilon^\mathrm{out}}}
\newcommand{\einput}{\epsilon_\mathrm{input}}
\newcommand{\echeck}{\epsilon_\mathrm{check}}
\newcommand{\ein}{{\epsilon_\mathrm{in}}}
\newcommand{\eout}{{\epsilon_\mathrm{out}}}
\newcommand{\nTT}{{n_T^\text{tot}}}
\newcommand{\nT}{{n_T}}
\newcommand{\nTb}{\overline{n_T}}
\newcommand{\cl}{{c_\mathrm{log}}}
\newcommand{\neo}{{c_\mathrm{out}}}
\newcommand{\nfin}{{n_\mathrm{lonely}}}
\newcommand{\nin}{{n_\mathrm{inner}}}
\newcommand{\kin}{{k_\mathrm{inner}}}

\newcommand{\nm}{{n_\mathrm{check}}}
\newtheorem{lemma}{Lemma}

\usepackage[normalem]{ulem}
\usepackage[usenames,dvipsnames,svgnames,table]{xcolor}

\begin{document}

\title{Magic State Distillation at Intermediate Size}

\author{Jeongwan Haah}
\affiliation{Quantum Architectures and Computation Group, Microsoft Research, Redmond, WA 98052, USA}

\author{Matthew B.~Hastings}

\affiliation{Station Q, Microsoft Research, Santa Barbara, CA 93106-6105, USA}
\affiliation{Quantum Architectures and Computation Group, Microsoft Research, Redmond, WA 98052, USA}

\author{D. Poulin}
\affiliation{D\'{e}partement de Physique \& Institut Quantique, Universit\'{e} de Sherbrooke, Quebec, Canada
}

\author{D. Wecker}
\affiliation{Quantum Architectures and Computation Group, Microsoft Research, Redmond, WA 98052, USA}

\begin{abstract}
Recently~\cite{hhpw} we proposed a family of magic state distillation protocols 
that obtains asymptotic performance that is conjectured to be optimal.
This family depends upon several codes, called ``inner codes'' and ``outer codes.''
In Ref.~\onlinecite{hhpw}, some small examples of these codes were given 
as well as an analysis of codes in the asymptotic limit.  
Here, we analyze such protocols in an intermediate size regime, using hundreds to thousands of qubits.
We use BCH inner codes~\cite{qbch}, 
combined with various outer codes.
We extend the protocols of Ref.~\onlinecite{hhpw} by adding error correction in some cases.
We present a variety of protocols in various input error regimes;
in many cases these protocols require significantly fewer input magic states 
to obtain a given output error than previous protocols.
\end{abstract}
%\date{\today}

\maketitle

\section{Introduction}
One widely-considered approach to building a fault tolerant quantum computer 
begins by implementing Clifford operations in a fault tolerant fashion, 
either using stabilizer 
codes~\cite{Gottesman1996Saturating,CalderbankRainsShorEtAl1997Quantum} 
or using Majorana fermions~\cite{karzig2017scalable}.
To obtain a universal quantum computer, 
it is necessary to supplement these operations with some non-Clifford operation.
A common approach is to use distillation of so-called magic states,
and inject them into quantum circuits.
In this method, many copies of a noisy magic state
are passed into a Clifford circuit, 
to create a smaller number of high quality magic 
states~\cite{Knill2004a,Knill2004b,BravyiKitaev2005Magic}.
The most common is to consider magic states 
that allow one to implement $\pi/4$-rotations ($T$-gates) by state injection,
but other methods are also considered.

Many distillation protocols have been 
considered~\cite{Knill2004a,BravyiKitaev2005Magic,MEK,BravyiHaah2012Magic}.
Recently, a family of protocols~\cite{hhpw} was suggested, 
where the number of input noisy magic states 
is conjectured to be asymptotically optimal 
($\gamma \rightarrow 1$ in the notation of \cite{BravyiHaah2012Magic}) 
with a relatively small space requirement.
The space overhead is important 
because some of the most efficient previously known protocols 
require overwhelmingly large numbers of qubits; 
we discuss this more quantitatively in Table~\ref{triocost} below,
giving space requirements for some protocols
based on concatenation of small distance codes, 
in some cases requiring in excess of $10^8$ qubits.

Ref.~\cite{hhpw} in fact is not a single protocol
but rather a recipe for building protocols.
It takes as ingredients several error correcting codes,
called inner codes and outer codes,
and from them defines a protocol.
In that paper, the asymptotic performance
$\gamma \rightarrow 1$ is achieved by combining various randomized and graph
theoretic constructions 
to show the existence of code families with the desired asymptotic properties.  
That paper also gave some small examples of the protocol 
using fewer than $30$ qubits, 
but these examples have notably worse performance 
than the asymptotic limit in terms of input $T$ count.

Thus, that paper~\cite{hhpw} left open an important question:
How well can this family of protocols do in an intermediate regime, 
using hundreds to thousands of qubits? 
In this paper, we begin to consider this question.
We present specific choices of inner and outer code 
which are useful in that regime.  
We also present several tricks,
which are useful to increase the performance of these protocols,
including concatenating with other protocols at various stages
and error correction.

We will give generalities to analyze the error rate of 
these distillation protocols
% in Section~\ref{enumerate} 
after setting up notation and convention 
in Section~\ref{bkgnd}.
The goal is not to prove a rigorous theorem of the form 
``the output error rate in this protocol is less than $\ldots$ for input error rate $\ldots$''  
Rather, the goal is to enumerate all error patterns 
at leading and next-to-leading order 
which give rise to logical errors.  
In later sections, we use this leading order enumeration 
to analyze specific protocols.  
We emphasize that the reader should be familiar with \cite{hhpw} to understand this paper.

\section{Background and convention}
\label{bkgnd}

We use the $T$-gate ($\pi/4$-rotation)
$T = e^{-i \pi Y/8} $
and employ a stochastic error model in which each 
$T$-gate has a $Y$ error with some probability $\ein$. 
We assume that the errors are independent between different $T$-gates.
Without loss of generality,
by a standard Clifford twirling argument, we can assume that 
each $\pi/4$ rotation and undistilled magic state 
suffers from independent $Y$ errors with probability $\epsilon$.
We refer to this error model as the ``stochastic error model".

Given a protocol and given an input error rate $\epsilon=\ein$, 
we define $\nbar$ to be the average number of output magic states.
In the basic family, the protocol either succeeds 
(and produces $\nout$ output magic states) 
or fails (and produces no output magic states) 
so that $\nbar$ is equal to the success probability times $\nout$.
In the generalizations of the basic family, 
we will in some cases discard some but not all of the magic states 
and output a number of magic states intermediate between $0$ and $\nout$.
In all of our protocols, the number of output magic states will always be at most $\nout$.

We define $\eout$ to be the probability that at least one of the output magic states has an error.  Note that if a protocol produces no output magic states (for example, if a protocol in the basic family fails), then by definition there are no output magic states with an error.
We define $\ebar$ by
\begin{align}
\ebar=\frac{\eout}{\nbar}.
\end{align}

This quantity $\ebar$ is the relevant measure of the output error rate 
if one uses a distillation protocol to produce $T$-gates to be used in a quantum algorithm with the following two properties.
First, the algorithm uses $\nTT$ total $T$-gates, with $\nTT \gg \nout$.
Second, no further error correction is used 
so that we are not willing to tolerate an error in any of the $T$-gates used.
Then, because of the first property, 
the number of times we need to call the distillation protocol 
is roughly $\nTT/\nbar$ and so 
the probability of an error in one or more output $T$-gates 
is roughly $\eout\nTT/\nbar=\ebar \nTT$ in the case that $\ebar \nTT \ll 1$.

A final relevant quantity for a protocol is $\nTb$, 
the average number of (approximate) $T$-gates used by the protocol 
(this includes both $T$-gates used inside an inner code, 
i.e., those $T$-gates used in performing a check of the outer code, 
as well as input magic states, 
where the input magic states refer to the $\nout$ approximate magic states which one produces at the start of the protocol).  
In general, the number of $T$-gates used by a protocol is a random variable which is why we must average: 
for example, in the basic family, 
if one detects an error partway through the protocol, one terminates without consuming any more $T$-gates, so that the number of $T$-gates used in that family is upper bounded by the number used if no errors occur.  For error corrected families, it is possible for more $T$-gates
to be consumed if errors occur.

Let $\nm$ denote the number of checks of the outer code.
Thus, for the basic family, the maximum number of $T$-gates used is $\nout+2\nm\nin$.

We will estimate $\nbar$, $\nTb$, and $\eout$ for the various families.  We do this by computing lower bounds on $\nbar$ and upper bounds on $\nTb$, and by expanding $\eout$ in a series in $\ein$ and computing upper bounds on the leading and next-to-leading order coefficients.  

We do these estimates by enumerating ``error patterns," 
where an error pattern refers to a particular choice of which input $T$-gates have an error.  
We will upper bound the number of error patterns which lead to an output magic state error and which have leading or next-to-leading weight.  Then, the leading order coefficient in $\eout$ is equal to the number of error patterns with leading order weight
and the next-to-leading order coefficient is upper bounded by the number of error patterns with next-to-leading order weight.
To see why this gives only an upper bound, imagine a simple toy model in which there are $3$ input $T$-gates and an error in one or more $T$-gates causes an output error.  Then, the probability of an output error is $1-(1-\ein)^3=3\ein-3\ein^2+\ein^3$.  However, there are $3$ error patterns with weight $1$ and $3$ error patterns with weight $2$ so that the enumeration correctly gives the leading order coefficient ($3=3$) but only upper bounds the next-to-leading order coefficient ($-3\leq 3$).  The reason for the discrepancy is simply that one ignores the factors of $1-\ein$ for the probability of no error on an input $T$-gate.
Later, we will more carefully considers these factors of $1-\ein$ to determine the next-to-leading order coefficient in $\eout$.

Finally, let us remark that while  we have used $\ein$ above to denote the input error probability, 
this error probability might not be the actual error probability of some physical gates.  Instead, one might wish to concatenate protocols, first reducing the error rate using some simple distillation protocol, and then using one of the protocols here.  The reason is that some of the protocols here use a larger number of $T$-gates and  so in order to have a reasonable chance of success it is necessary for the input error rate to be not too large.  When we give specific numbers later, we will pick $\ein=10^{-3}$ in almost all cases, but one should assume that this number may be the result of the output of some other simple distillation protocol. 
For example, a single round of the  Meier-Eastin-Knill protocol~\cite{MEK} (called ``MEK" below) will produce an output error rate slightly smaller than $10^{-3}$ from an input error rate of $10^{-2}$.

In fact, in some cases, we will prefer to use the output of one distillation protocol to produce the input magic states and use the output of another distillation protocol to produce the approximate $T$-gates used inside an inner code.  To analyze this situation, we need to give each error pattern two distinct weights: we say that an error pattern has weight $(a,b)$ if it has $a$ input errors and $b$ errors inside $T$-gates.
We generalize the notion of an input error rate to define two types of input error rate; let $\einput$ be the error probability on the input magic states and let $\echeck$ be the error probability on the
$T$-gates inside the inner code.  When we give specific numbers later, 
we will set $\echeck=\ein=10^{-3}$, 
but in some cases we will assume that additional concatenation is employed 
so that $\einput \ll \echeck$;
in this case, we will take for simplicity $\einput=9 \times 10^{-6}$ 
as this is roughly the error after the MEK protocol is applied to an input error of $10^{-3}$.

When expanding in both $\echeck,\einput$,
we will describe a term as being at order $\ein^k$ to indicate that the total power in $\echeck,\einput$ is equal to $k$. 

Let the leading order patterns have weight $d$.
We let $c_{a,b}$ denote the number of error patterns with weight $(a,b)$ that lead to an output error
so that
\begin{align}
\eout \leq \sum_{a+b=d} c_{a,b} \einput^a \echeck^b + \sum_{a+b=d+1} c_{a,b} \einput^{a} \echeck^{b} + O(\ein^{d+2}).
\end{align}
Hence, the inner and outer codes must both have distance $\geq d$.
In addition, any $k$ qubit error in the outer code, for $k < d$, must violate at least
\[
     \left\lceil \frac{d-k}{2} \right\rceil 
\]
checks of the outer code.
We refer to these assumptions as the ``distance and sensitivity bounds".
We consider normal inner codes throughout, on which
the transversal Hadamard is the simultaneous logical Hadamard on all logical qubits.

We use the term ``inner code syndrome'' to refer to a measurement of an inner code stabilizer that shows an error (i.e., a nontrivial nontrivial inner syndrome) 
and ``outer code syndrome'' to refer to a measurement of an outer code check that shows an error (nontrivial outer syndrome).
We use the term ``incorrect outer code measurement'' 
to indicate a measurement of the outer code that differs from 
the correct value even though no inner code syndrome occurs; 
this requires at least $2$ errors inside the inner code.
To measure an outer code check, 
the measurement is implemented using an inner code,
and the outer code check is read by measuring an ancilla bit; 
if a stabilizer measurement of the inner code shows an error, 
we ignore the value of the ancilla bit.
That is, the outer code measurement is only meaningful 
if there is no inner code syndrome.

We use a set notation to define checks of the outer code,
saying $q \in C$ to indicate that a qubit $q$ is in a check $C$.
That is, an outer code check measures the eigenvalue of 
$\bigotimes_{q \in C} H_q $ where $H_q$ is the Hadamard operator on qubit $q$. 

We define a bipartite graph, called the Tanner graph, 
where one set of vertices corresponds to checks and the other set corresponds to qubits,
such that a qubit is in a check if there is an edge between the corresponding vertices.
This graph will be useful in applying graph-theoretical ideas to the analysis of the outer code.

\section{Basic Family}
\label{basic}

In this section, 
% we define various families of distillation protocols 
% and give general results on analyzing the error rate in these protocols 
% by enumerating different error patterns which give rise to an output error.  
% We begin with some background and definition of terms in \ref{bkgnd}.  
% We then enumerate errors in various various families of distillation protocols.
% In \ref~{basic}, (JH: this is said in the introduction.)
we consider a family of protocols that is defined in Ref.~\cite{hhpw}
with the outer code having $\nout$ qubits.
A protocol in this family is defined by a weakly self-dual CSS code 
$[[\nin,\kin,d]]$ (inner code),
which is used to implement controlled $H^{\otimes \kin}$,
and a parity check matrix (outer code),
which specifies which and when input magic states are tested.
We call this family the {\it basic family}.

The other families below 
will be generalizations of the basic family
by employing error correction, rather than mere detection.
These generalizations will have a higher output error rate,
but they will be more likely to suceed in producing an output magic state;
in the basic family, if an inner code stabilizer measurement 
shows an error or an outer code check shows and error, 
the protocol is terminated, giving a failure with no output magic states, 
while in the generalizations we sometimes attempt to correct errors.

Some of our estimates in the enumeration will be conservative, 
in that they will upper bound the number of error patterns 
which lead to an output error.
This is partly done for generality of the results.
We will comment on this later when it becomes important.

\subsection{Output count}

In any of the families of protocols that we consider, 
including the basic family in particular, 
if there are no input $T$-gate errors, 
then the protocol succeeds and output $\nout$ magic states.  Hence,
\begin{align}
\label{nbarlb}
\nbar \geq \nout (1-\einput)^{\nout}(1-\echeck)^{2\nm\nin}.
\end{align}
Since there are some error patterns with nonzero weight that do not cause an inner code or outer code syndrome, Eq.~(\ref{nbarlb}) indeed is only a lower bound.  However, for the basic family, Eq.~(\ref{nbarlb}) is quite accurate; the families considered later will include error correction that increases $\nbar$.

The quantity $\nTb$ is upper bounded by the number of 
$T$-gates used if no errors occur.
Let $\nm$ denote the number of check operators in the outer code; thus, $\nm$ is equal to the number of times that one encodes qubits into the inner code.
Let $\nin$ denote the number of physical qubits in the inner code.  Hence,
\begin{align}
\nTb \leq \nout+ 2 \nm \nin.
\end{align}

This number is only an upper bound, 
because if we terminate the protocol partway through, no further $T$-gates will be consumed.
Suppose, for example, that we execute the inner code checks sequentially 
and if a single inner code check fails, 
then no further checks are executed 
(this will not always apply since in some cases we may choose to execute checks in parallel).
Suppose also that $2\nin \echeck \ll 1$.
Then, the most likely way in which an inner code syndrome can occur is 
for there to be a single error inside the inner code. 
The probability that no inner code syndromes occur in the first $j$ checks 
is roughly $(1-2\nin \echeck)^j$.  Then,
\begin{align}
\nTb \lesssim \nout +  2\nin \sum_{j=0}^{\nm-1} (1-2\nin\echeck)^j. \label{withterminate}
\end{align}
This estimate itself is still an overestimate 
as it does not include the possibility of terminating the protocol early 
due to outer code syndromes.

\subsection{Error Patterns With a Logical Error}
\label{epwalp}
We now enumerate error patterns such that some logical error occurs in an inner code.

Let $\cl(w)$ denote the number of nontrivial logical operators of the inner code with weight $w$, 
where these logical operators are products of Pauli $Y$ operators. 
Hence, $\cl(w)=0$ for $w<d$.
One possible error pattern leading to an output error 
is that in some inner code, 
errors in $T$-gates produce a nontrivial logical operator.
The numbers of such patterns of weights $d$ and $d+1$ are bounded by
\[
    \nm 2^d \cl(d)\; , \quad \nm 2^{d+1} \cl(d+1),
\]
respectively.

The factors of $2^d$ and $2^{d+1}$ arise for the following reason: 
$\cl(w)$ is the number of logical error operators of weight $w$ in the inner code.  
Each such operator is a product of Pauli $Y$ operators on certain physical qubits of the inner code.  
Hence, we can produce such a logical error by errors in the $T$-gates on those qubits.  
However, there are two $T$-gates acting on each physical qubit of the inner code.
This gives the factor of $2^w$.  

Note that for some such error patterns, we may encounter an outer code syndrome.  
This possibility of an outer code syndrome will reduce the number of error patterns which lead to an output error.
For example, consider an error pattern $P$ in which all of the errors occur on the first $T$-gate in the pair of $T$-gates on a qubit.  This error pattern is equivalent to applying some logical operator $L$ and then measuring the outer code check given that that logical operator is applied.  If there is an outer code syndrome for this check, then no output error occurs.  On the other hand, modify pattern $P$ to define a new pattern $P'$ in which one of the errors in $P$ is moved from the first $T$-gate in a pair to the second $T$-gate in a pair.  Then, this error pattern is equivalent to applying logical operator $L$ followed by incorrectly measuring the given outer code check.  Hence, if pattern $P$ does not lead to an outer code syndrome, then pattern $P'$ will.  For this reason, the number of patterns that lead to an output error is only half that given above:
\[
\nm 2^{d-1} \cl(d)\; , \; \nm 2^{d} \cl(d+1),
\]
respectively.

In fact, these numbers are overestimates of the number of patterns producing an output error.
Suppose that no $T$-gates have an error, except for some number of $T$-gates inside 
an inner code leading to a logical error. 
Suppose this inner code was used to implement some outer code check $C$.
Thus, after this particular step, 
at least one of the qubits checked by the outer code check $C$ has an error.
If every qubit $q \in C$ is subsequently checked 
by at least one outer code check $C'$ such that 
$C'$ does not contain any other qubit $q' \neq q$ with $q' \in C$ 
(i.e., $\forall q \in C~ \exists C' \; {\rm subsequent} \; {\rm to} \; C ~:~ C \cap C' = \{q\}$),
then some outer code check $C'$ will detect an error up to subleading terms in $\echeck$ 
(with probability $O(\echeck^2)$, the outer code check $C'$ detects no error even if there is one).
So, let $\nfin$ denote the number of ``{\bf lonely checks}''; 
these are measurements $C$ of the outer code 
which do {\em not} obey the property that 
every qubit $q\in C$ is subsequently checked by some check $C'$ 
such that $C'$ contains only one qubits from $C$.
So, the number of such pattern of weights $d$ and $d+1$ are bounded by
\begin{align}
\nfin 2^{d-1} \cl(d)\; , \; \nfin 2^d \cl(d+1),
\end{align}
respectively.

Another possible pattern to consider is that 
in some inner code there are exactly $d$ errors which produce a logical error 
and that there is exactly one other $T$-gate error. 
The total weight of this error pattern is $d+1$.  
Let us first emphasize that error patterns of this nature do indeed exist 
such that no inner code syndrome or outer code syndrome occurs.
For example, an erroneous input magic state (one error) that is to be checked
gets corrected by a logical error from some inner code ($d$ errors).
However, while this error pattern leads to no inner code syndrome or outer code syndrome,
it also does not lead to an output magic state error.

So, we next consider whether there are any error patterns of weight $d+1$ which
include an inner code logical error of weight $d$ during implementation of
an outer code check $C$ and include one other $T$-gate error $E$, 
which overall produces an output error.
We claim that if the first check on each qubit is not a lonely check,
then no such pattern exists.
The $T$-gate error $E$ must be in one of the input magic states, say on $q$;
otherwise, it will lead to an inner code syndrome.
If $q$ is checked by some check $C'$ before $q$ is checked by $C$, or if $q$ is not in $C$, 
then at leading order in $\echeck$, 
the check $C'$ will give an outer code syndrome.
Hence, the first check on $q$ must be $C$.
However, since $C$ is not lonely by assumption,
some subsequent check must give an error.

\subsection{Error Patterns With No Logical Error}
Having enumerated error patterns where some logical error occurs in an inner code,
we now consider error patterns where no logical error occurs.  
Thus, the output state is the same as the input state, 
and the only way an output error can occur 
is for some number of input states to be incorrect 
and then for some number of outer code checks to be measured incorrectly.
Let $\neo(u,v)$ denote the number of bit patterns on $\nout$ bits 
with Hamming weight $u$ that violate $v$ checks of the outer code.
By assumption,
\begin{align}
u+2v < d \quad \Longrightarrow \quad
\neo(u,v)=0.
\end{align}

Then, for $a>0$ and $b$ even, 
the number of $T$-gate error patterns of weight $(a,b)$ 
with no inner code logical error and 
with no inner code syndrome or outer code syndrome
and with an output magic state error 
is equal to
\begin{align}
\sum_{j=0}^{b/2} \neo(a,j) 
\sum_{\substack{f_i > 0 \\ f_i \text{odd}\\ \sum_{i=1}^j f_i = b/2 }} \prod_{i=1}^{j} \binom{\nin}{f_i}
\end{align}
The combinatorial factor arises 
because each incorrect outer code measurement 
can be due to an odd number of pairs of $T$-gate errors 
on any of the $\nin$ different qubits in that inner code.

\subsection{Summary}
Combining the above results, we find that:
\begin{lemma}
\label{basiclemma}
Let $c_{a,b}$ be the number of error patterns that lead to an output error
where $a$ is the number of input errors, 
and $b$ the number of errors inside the inner codes.
Let $\neo(u,v)$ be the number of input error patterns
of weight $u$ that violate $v$ checks of the outer code.
Let $\cl(w)$ be the number of $Y$-logical operators of weight $w$ of the inner code.
Using the basic protocol, 
assuming that the codes obey the distance and sensitivity bounds 
and that the first check on each qubit $q$ is not a lonely check, 
we have for $a+b=d$ or $a+b=d+1$
\begin{align} 
 c_{0,b} &\leq \nfin 2^{b-1} \cl(b) & (a=0)\\
 c_{a,b} &= 
\sum_{j=0}^{b/2} \neo(a,j) 
\sum_{\substack{f_i > 0 \\ f_i ~\mathrm{odd}\\ \sum_{i=1}^j f_i = b/2 }} \prod_{i=1}^{j} \binom{\nin}{f_i} & (a>0).
\end{align}
When $ a+b \le 6$ the second formula becomes 
\begin{align}
   c_{a,b} = \nin^{b/2}\neo(a,b/2) .
\end{align}
\end{lemma}

\section{Inner Code Corrected Family}
\label{icc}
We now discuss a modification to the basic family to increase the probability that the protocol succeeds.
This modification involves error correcting the inner code.  
We define an integer called the ``order" of the error correction; the order $0$ inner code corrected family will be the same as the basic family.
As before, we will assume that the outer code obeys the distance and sensitivity properties and we will assume that the first check on any qubit is not a lonely check.

The order $m$ inner code corrected family modifies the basic protocol as follows.
Whenever we perform an outer code check, we use the following algorithm (some of the terms in this algorithm are explained below):
\begin{itemize}
\item[1.] Measure the outer code check using the inner code.  If no inner code syndrome occurs but an outer code syndrome occurs, terminate the protocol.  If no inner code syndrome occurs and no outer code syndrome occurs, continue the protocol; i.e., we are done with measuring this outer code check and we proceed to the next one.  
When done measuring all checks, go to step {\bf 3}.
Otherwise, if an inner code syndrome occurs, go to step {\bf 2}.

\item[2.] If the observed error syndrome for the inner code can be produced by an error pattern of weight $w \leq m$ then error correct (see below) and repeat step {\bf 1}.  If the observed error syndrome cannot be produced by an error pattern of weight $w\leq m$, terminate the protocol.

\item[3.] When the protocol is done, we discard any qubits 
that might be ``lower quality,'' 
as explained below, before outputting the other qubits.  
Hence, the number of magic states produced may be smaller than $\nout$.
\end{itemize}
Note that steps $1,2$ can be repeated an arbitrary number of times 
for a given check.

Applying error correction, can be performed as follows.  
% Let $w$ be a minimum weight error pattern which produces the observed inner code syndrome.  
The decoding circuit of the inner code is some Clifford operator, 
as is the encoding circuit that is the inverse of the decoding circuit.
The decoding circuit maps $\nin$ physical qubits to $\kin$ logical qubits and $\nin-\kin$ ancilla qubits.  
Measuring the inner code stabilizers of a state is accomplished by measuring these ancillas.
To error correct,
identify, by classical computation,
a product of $Y$ on the inner code of minimum weight 
that matches the observed syndrome,
compute the affected logical qubits,
and then apply $Y$ on those affected logical qubits.
This is slightly different from usual error correction
where syndrome extraction and correction are performed without a decoding circuit,
but the procedure here is more efficient for our purpose
because a nontrivial syndrome is not too frequent
and a different set of logical qubits will be subsequently
encoded into a different outer code check.
% if some of the ancillas have errors, first apply the encoding circuit to the state which has error on the ancilla qubits.  The resulting state will not be a code word.  Then apply a product of Pauli $Y$-operators corresponding to the bits in error pattern $w$.  Finally, applying the decoding circuit.  In fact, one can simplify this procedure: since the encoding and decoding circuit are inverses of each other and are Cliffords, the effect is the same as applying some product of Pauli operators to the state of $\kin$ logical qubits and $\nin-\kin$ ancilla qubits.

The concept of ``{\bf lower quality}'' at the end of the protocol is as follows.
We say that a qubit $q$ is lower quality 
if $q$ belongs to an outer code check $C$ such that 
when $C$ is measured, an inner code syndrome occurs that is corrected 
(i.e, case $2$ occurs one or more times when measuring $C$) 
and if there is no other check $C'$ after $C$ 
such that $C' \cap C = \{ q \}$, a singleton.
Note that this means $C$ must be a lonely check.
Generally, error correction can lower the fidelity of the output state
because an error pattern of weight $d-1$ may be confused with a weight $1$ error.
By discarding lower quality qubits,
we eliminate this possibility
that a qubit is contaminated by error correction
as the qubit is tested once more after the error correction.
This allows us to maintain the order of reduction in error at $d$,
even with error correction of order $m=1$. 

At a higher order of error correction,
it is necessary to broaden the class of ``lower quality'' qubits.
In this modification,
we say that a qubit $q$ is lower quality 
if $q \in C$ for some outer code check $C$ such that when $C$ is measured, 
an inner code syndrome occurs that is corrected 
(i.e., case $2$ occurs one or more times when measuring that check) 
and if there is {\it at most one} other check $C'$ after $C$ 
such that $C' \cap C = \{q\}$.
We call a protocol ``conservative'' 
if this broadened class of lower quality qubits are discarded.
The conservative protocol should be used 
when error correction of order $m=2$ is employed.
Unless otherwise specified, we do not use the conservative protocol.

We will assume throughout that $m \leq 2$. 
It will become clear at the end of the next subsection 
why there is little reason to consider protocols with $m \geq 3$.
We also assume that the inner code has a distance $d\geq 5$.
Note that this means that $2(d-m)>d$.

\subsection{Estimates for $\nbar$ and $\nTb$}

To estimate $\nbar$, 
note first that the probability that none of the input magic states 
has an error is equal to $(1-\einput)^{\nout}$.
Assume that none of the input magic states has an error.
For each measurement of an outer code check, 
one of several possibilities can occur:
\begin{itemize}
\item[A.] No $T$-gate errors occur.  This happens with probability $(1-\echeck)^{2\nin}$.

\item[B.] An inner code syndrome occurs 
and this error can be corrected by an error pattern of weight $w \leq m$.  
This occurs with probability at most $2\nin \echeck + O(\echeck^2)$ for $m=1$ and
$2\nin \echeck +  4 \nin^2 \echeck^2+O(\echeck^3)$ for $m=2$.  
To see this, note that a single $T$-gate error can occur in any of $2 \nin$ locations, 
and such an error pattern occurs with probability 
$2 \nin \echeck (1-\echeck)^{2\nin-1} \leq 2 \nin \echeck$. 
A pair of $T$-gate errors can each occur in any of $2\nin$ locations 
so that the probability is at most 
$\binom{2 \nin}{2} \echeck^2 (1-\echeck)^{2\nin-2} \leq 4 \nin^2 \echeck^2$.  
In fact, the probability is slightly smaller than this 
because if the two $T$-gate errors occur on the same qubit, 
then no inner code syndrome occurs, 
but an outer code measurement error occurs.

\item[C.] An outer code measurement error occurs due to an incorrect outer code measurement.  
This happens with probability at most $\nin \echeck^2 + O(\echeck^4)$.

\item[D.] An inner code syndrome occurs that cannot be corrected by an error pattern of weight $w \leq m$.  
This occurs with probability at most $(2\nin)^{m+1} \echeck^{m+1}/(m+1)! + O(\echeck^{m+2})$.

\item[E.] A logical error occurs in the inner code with no inner code syndrome.  
This occurs with probability $O(\echeck^d)$.
\end{itemize}
In case A, the measurement of the inner code is done.  
In case B, we repeat the measurement.
In cases C and D, we terminate the protocol.
In case E, we do not immediately terminate the protocol, 
but it may be terminated subsequently.  

As a lower bound on the probability that the protocol does not terminate, 
let us assume that case E does not occur.
We have a Markov chain each time we measure a check: 
With probability  
\[
P_{succ}\equiv (1-\echeck)^{2\nin},
\]
no error occurs, and we are done with the measurement and proceed to the next measurement.
With probability 
\begin{align*}
P_{repeat} \equiv 
\begin{cases}
2\nin \echeck +O(\echeck^2) & \text{if } m =1,\\
2\nin \echeck +  4 \nin^2 \echeck^2+O(\echeck^3) & \text{if } m = 2,
\end{cases}
\end{align*}
we repeat the measurement.
With probability at most
\[
P_{fail} \equiv \nin \echeck^2+ (2\nin)^{m+1} \echeck^{m+1}/(m+1)!+O(\echeck^{m+2}),
\]
the protocol is terminated with failure.

Thus, one can straightforwardly calculate that 
the probability that we proceed to the next measurement is 
$P_{succ}/(P_{succ}+P_{fail})$, 
while the probability that the protocol
is terminated with failure on any given measurement is 
$P_{fail}/(P_{succ}+P_{fail})$.
Hence, the probability that the protocol succeeds is lower bounded by
\begin{align}
(1-\einput)^{\nout} \Bigl(\frac{P_{succ}}{P_{succ}+P_{fail}}\Bigr)^\nm.
\end{align}
Given that the protocol succeeds, 
the expected number of qubits that are lower quality can be bounded as follows.
For each lonely check, the probability that case B occurs is $P_{repeat}$.
Hence, the average number of qubits that are lower quality is at most $\nfin P_{repeat} \kin$,
and so
\begin{align}
\nbar \geq \Bigl(\nout-\nfin P_{repeat} k_{in}\Bigr) (1-\einput)^{\nout} 
           \Bigl(\frac{P_{succ}}{P_{succ}+P_{fail}}\Bigr)^{\nm}.
\end{align}

Now we can see why there is little reason to consider $m\geq 3$. 
The probability of failing due to an error pattern with weight $w>m$ (i.e., case D above) is $O(\echeck^{m+1})$.
As we increase $m$, we reduce the probability of failing due to such an error pattern.   
However, the probability of failing due to an outer code measurement failure (case C above) is of order $\echeck^2$.
So, once we have sufficiently large $m$ that the probability of case C is much larger than that of case D, 
there is little reason to consider larger $m$.  
At $m=1$, case C and case D are both at the same order in $\echeck$ but case D has a much larger prefactor, 
so there is some reason to consider  $m=2$.  
At $m=2$, case D is higher order in $\echeck$ than case C and so there is little reason to consider $m\geq 3$.

Finally, the average number of times that we repeat a measurement is 
$1+P_{repeat}+P_{repeat}^2+\cdots = 1/(1-P_{repeat})$.
Hence,
\begin{align}
\label{withouttermerrc}
\nTb \leq \nout + \frac{2 \nin \nm}{1-P_{repeat}}.
\end{align}
In fact, following similar reasoning as lead to Eq.~(\ref{withterminate}), 
we can reduce this estimate for $\nTb$ 
if we terminate the protocol as soon as an error is detected.
In this case, we find that
\begin{align}
\label{withtermerrc}
\nTb 
\leq 
\nout + \frac{2 \nin}{1-P_{repeat}} \sum_{j=0}^{\nm-1} \Bigl( \frac{P_{succ}}{P_{succ}+P_{fail}}\Bigr)^{j}
\end{align}

\subsection{Estimates for $\eout$}
We now estimate $\eout$.  
We will enumerate three different types of error patterns,
depending on whether case B (error correction) or a logical error occurs.
We will assume that every qubit $q$ is in at least two checks of the outer code; 
this is necessary to obtain fifth or higher order reduction in error.
Also, we only consider ``$4$-cycle free'' outer codes
where no pair of checks $C, C'$ share more than one qubit; 
i.e., $|C \cap C'| \le 1$ for any distinct checks $C, C'$ (we call these $4$-cycle free because it implies that the Tanner graph has no $4$-cycles).

There are three cases to consider:
(i) No inner code syndrome ($c^0_{a,b}$),
(ii) some inner code syndrome that get removed by error correction ($c^1_{a,b}$),
and (iii) some inner code syndrome that becomes a logical error by error correction ($c^{log}_{a,b}$).
Then, $c$ will be the sum of $c^0,c^1,c^{log}$.

\subsubsection{Logical error without inner code syndrome}

First, let us consider error patterns in which no inner code syndrome occurs; 
cases A, C, E occur only.
In this case, we can use the analysis that led to Lemma~\ref{basiclemma}
to show that the number of error patterns with total weight $w=d$ or $w=d+1$ leading to an output error 
is bounded by
\begin{align}
c^0_{0,b} &\leq \nfin 2^{w-1} \cl(w) &\text{(no input $T$-state error)}, \\
c^0_{a,b} &=\nin^{b/2} \neo(a,b/2) & \text{(error on $a>0$ input $T$-states)},
\end{align}
for $a+b=d \le 6$ or $a+b=d+1 \le 6$.

\subsubsection{Error patterns with Inner Code Error But No Logical Error from Correction}

We now consider the case that there is an inner code syndrome, 
i.e., case B occurs at least once.
In this subsubsection, we analyze the case that no logical error occurs due to our correction, while in the next subsubsection we consider the case of a logical error due to correction.

Suppose that a weight 1 error occurs inside the inner code of a check $C$,
which is corrected without logical error.  All such error patterns of total weight $d+1$ leading to an output error can be constructed in the following way: consider an error pattern
$P$ of total weight $d$ that leads to an output error.
Let $C$ be any check. Define a new pattern $P'$ where $P'$ has the same error pattern as $P$ except that when one first measures check $C$, first a  single $T$-gate error occurs inside the inner code, then one corrects this error, and then one continues as in error pattern $P$.  These patterns $P'$ are the patterns that we consider in this subsubsection.

First consider the case that
that $P$ has a logical error on check $C$.
If $C$ is lonely, the affected logical qubit will be discarded,
and if $C$ is not lonely, the logical error will be detected,
so this pattern does not lead to an output error.
Thus, there are no such patterns of total weight $d+1$ giving an output error.

Now consider other choices of $P$ and $C$, so that $P$ does not have a logical error on check $C$.  For a given
pattern $P$, there are either $2\nm \nin$ or $2 (\nm-1) \nin$ ways to construct such a pattern (there are $2\nin$ places to insert  weight $1$ error inside a check, and there are $\nm$ checks, though the check $C$ cannot be on a check where $P$ has a logical error so there may be only $\nm-1$ such checks).
The sum of such patterns
may be much larger than other terms of total weight $d+1$ 
due to the large factor $2 \nm \nin$,
but its contribution to the output error {\it probability} is not too large.
Recall that when we compute the probability of an output error, 
this probability is {\it not} obtained simply by 
$\sum_{a,b} \einput^a \echeck^b c_{a,b}$ 
but rather one must also include the probability that other gates do not have an error.
For example, in the basic protocol (where the total number of $T$-gates is gixed), 
we must instead compute $\sum_{a,b} \einput^a (1-\einput)^{\nout-a} \echeck^b (1-\echeck)^{2\nin\nm-b} c_{a,b}$.  When we include this probability that other gates do not have an error,
our error probability at order $\ein^d$ is indeed given by $\sum_{a+b=d} \einput^a \echeck^b c_{a,b}$, 
but our error probability at order $\ein^{d+1}$ is given by 
\begin{align}
\sum_{a+b=d+1} \einput^a \echeck^b c_{a,b}
-\sum_{a+b=d} \einput^a \echeck^{b+1} (2\nin \nm-b) c_{a,b}.
\end{align}
This term $-\sum_{a+b=d} \einput^a \echeck^{b+1} (2\nin \nm-b) c_{a,b}$
will largely cancel a term $c^1_{a,b} = c^0_{a,b-1} 2\nm\nin$ up to terms 
$-\sum_{a+b=d} \einput^a \echeck^b b c_{a,b}$.
Thus, while the series expansion for the total number of errors of a given weight 
has this large prefactor $\nm \nin$ at order $d+1$ 
which may, in many cases, make the order $d+1$ term comparable to the order $d$ term,
the series expansion for the output error probability has these terms largely cancel at order $d+1$.
One may indeed show a formal cancellation of these terms at higher orders;
however since our goal in this paper is not a formal proof of the error probability,
but rather we are content with an accurate estimate, 
we will simply estimate the error probability at order $d,d+1$ 
and use this cancellation at that order.

\subsubsection{Error patterns with Inner Code Error leading to a Logical Error by Correction}

Now suppose a logical error occurs after error correction in some check $C$.  Hence, an error pattern of weight at least $d-m$ occurred inside that inner code such that after applying an additional weight error correction of weight at most $m$, a logical operator of weight at least $d$ was produced. Note that if the error pattern, including errors both inside the inner code measuring $C$ and other errors elsewhere, has weight at most $d$, then such a logical error can occur only once as $2(d-m)>d$.  Even if we also wish to consider error patterns with weight at most $d+1$, then so long as $2(d-m)>d+1$, such a logical error can also occur only once.
Let us restrict then to the case where such a logical error can occur only once.

It will be easier to give bounds on the number of error patterns which can cause an output error for specific outer codes, rather than general bounds, so let us first describe the general procedure and then give a few bounds which hold in generality.  The general procedure is to consider all input states to the outer code, with some given number of errors with total weight $w_{in}$, then consider all places at which a logical error can occur.  Such a logical error requires at least $d-m$ additional errors and one must sum over the different locations in which they occur.  Then, for each such place where a logical error can occur, one must consider all possible states of the qubits after the logical error (i.e., they are in some initial state before the check and some state after).   One must then count the number of violated checks; each violated check requires an additional two errors but gives an additional prefactor equal to $\nin$ describing the number of places these errors can occur.  The total number of violated checks due to input errors, logical error, and measurement errors, must be at most $d$ or $d+1$.

This general procedure includes a sum over error patterns inside the inner code leading to a logical error.
The number of error patterns leading to a given logical error can be computed for any given inner code.  As an upper
bound, the number of error patterns leading to given logical error is upper bounded by the number of error patterns
leading to an arbitrary logical error.
The number of weight $d-2$ errors in a check $C$ 
that lead to a logical error after error correction in the case $m=2$ can be estimated as follows.
We have defined $\cl(d)$ as the number of nontrivial logical operators of weight $d$.
Each error pattern of weight $d-2$ that leads to an error after error correction 
is given by removing $2$ errors from such a logical operator.
Hence, there are at most
\[
2^{d-2} \binom{d}{2}  \cl(d)
\]
such patterns.
Similarly, the number of weight $d-1$ errors that lead to a logical error is bounded by 
\[
2^{d-1} d \cl(d).
\]

We now give some general results on this summation.

Suppose there was no error on qubits of a check $C$ before we measure $C$,
but after we measure $C$, a qubit $q \in C$ becomes erroneous.
Any qubit $q$ which is not lower quality 
will be in at least one more check $C'$ 
which shares exactly one qubit (namely, $q$) with $C$.
If all other qubits in $C'$ have no error,
then in order for no outer error to occur when measuring $C'$,
there must be an incorrect outer code measurement.
Then, for $m=1$ there are no such error patterns with total weight $d$,
but the number of such error patterns of total weight $d+1$ is
bounded by $d \cl(d) \nin n_{once}$,
where $n_{once}$ is the number of possible checks $C$ such that
only one check $C'$ after $C$ will be violated for some logical error in $C$.
For $m=2$, the number of such error patterns of total weight $d$ is bounded by 
$\binom{d}{2} \cl(d) \nin n_{once}$,
and that of total weight $d+1$ is bounded by $d \cl(d) \nin n_{once}$.
If we use the conservative protocol,
qubits that are not lower quality must be in at least two more such checks $C'$,
and so there are not such error patterns of total weight $d$ or $d+1$.

Suppose instead that there were some errors in qubits in $C$ before the error correction in $C$,
but after the error correction, none of the qubits in $C$ have an error.
So, the result of the error corrections 
was to turn a state with an error on one or more qubits in $C$ 
into a state with no errors on qubits in $C$.
We claim that in this case,
there are no error patterns of weight $d$ or $d+1$
leading to an output error, even for $m=2$, whenever the outer code is $4$-cycle free.
For an output error to occur,
there must have been an error on at least one input qubit $q'$ with $q' \not \in C$.
This error is in addition to an error on $q \in C$ before the error correction in $C$.
The logical error in $C$ needs at least $d-2$ errors (or $d-1$ if $m=1$).
So, to have an error of total weight $d$ or $d+1$,
we are left with at most 1 more position to put an error on.
Since we are assuming at least two checks per qubit,
some check on $q'$ must have an incorrect outer code measurement,
which requires at least two more erroneous $T$ gates,
and hence the total weight of error would exceed $d+1$.

Suppose instead that there are errors in qubits in $C$ both before and after the logical error.  
We will now show that in this case also, 
there are no error patterns of weight $d$ or $d+1$ 
leading to an output logical error, even for $m=2$.
By the analysis of the above paragraph,
for a $4$-cycle free outer code 
we can assume that the qubits in the complement of $C$ do not have an error
if the total error pattern has weight $d$ or $d+1$.
If two qubits in $C$ have an input error, 
then, in order for the total error pattern to have weight $d$ or $d+1$,
$d-2$ or $d-1$ $T$ gates in $C$ must be erroneous,
and then there is no room for incorrect outer code measurements to occur.
But then, the qubits in $C$ will be lower quality and discarded.
If only one qubit in $C$ has an input error,
then no error correction (case B) should be performed (by the preceding argument),
and a logical error of weight $d$ should occur.
Since any qubit is in at least two checks, 
a check either before or after $C$ must reject the input.
% Therefore, no such error patterns can occur with total weight $d$ or $d+1$ for a $4$-cycle free outer code.

To summarize, enumerating error patterns involving logical errors,
we find
\begin{align}
c^{log}_{a,b}&=0 \text{ where } a+b=d,\\
c^{log}_{0,d+1} &\leq 2^{d-1} d \cl(d) \nin n_{once} & (m =1),\nonumber
\end{align}
while
\begin{align}
c^{log}_{0,d}    &\le 2^{d-2} \binom{d}{2} \cl(d) \nin n_{once}, \\
c^{log}_{0,d+1} &\le 2^{d-1} d \cl(d) \nin n_{once} & (m=2), \nonumber
\end{align}
where $n_{once}$ is the number of possible checks $C$ such that 
only one check $C'$ after $C$ will be violated for some logical error in $C$.

Note that the bound is an overestimate, 
and in some cases the identified source of error in this subsection 
may be eliminated completely
by a choice of logical operators.
Indeed, if, under a certain choice, 
any logical operator of weight $d$ always affects two or more logical qubits 
in a check that is not lonely,
then such an error pattern requires two or more incorrect outer code measurements.
Thus, whenever possible, it is better to choose logical operators of earlier checks
such that small weight logical operators act on many logical qubits.

\section{Further variants}

\subsection{Inner and Outer Code Corrected Family}
\label{iccocc}
A final generalization that one might consider 
is a family using both inner and outer code error correction.
Thus, one might perform inner code error correction as before, 
but if some outer code syndrome occurs, 
one attempts to either error correct, or one discards certain states 
rather than terminating the protocol.
We will not consider this possibility further in this paper,
beyond mentioning that it is possible.
The reason is as follows: 
the primary goal to consider outer code error correction 
is to reduce the probability of terminating the protocol 
if there is an outer code syndrome due to an input magic state error.
(Terminating the protocol due to the case 
where there is no input magic state error 
but there is an incorrect outer code measurement is much less likely.)
By using concatenation to reduce $\einput \ll \echeck$,
we can make this probability quite small.
For the families of protocols that we consider,
the quantity $\nout$ is fairly small compared to the total number of $T$-gates $\nT$,
so that there is little cost in doing this additional concatenation.
However, for even larger protocols it may be worth considering outer code error correction.

\subsection{Partial Restart}
\label{pr}
A further modification of the protocol is what we call a ``{\bf partial restart}.''
Suppose that in some protocol, we start by measuring some checks 
$C_1,C_2,\ldots,C_k$ for some $k$, 
such that $C_i \cap C_j =\emptyset$ for $1\leq i < j \leq k$.
The protocol will involve later measuring additional checks;
we are simply describing a set of checks that can be measured in parallel 
at the start of the protocol.
(The grid code below gives an example of such a protocol,
where all the ``vertical" checks can be measured in parallel before all other checks; concatenated codes provide another familiar example of this, in that when a low-level block in a concatenated code fails, only that block is discarded).
Suppose that one or more of the checks gives an outer code syndrome.
The protocol explained above then involves terminating the protocol with failure,
discarding all qubits and restarting.
However, in fact it is only necessary to discard and re-measure 
the qubits in the checks that have an outer code syndrome.
That is, if check $C_1$ gives an outer code syndrome, 
but none of the others do, we can re-prepare approximate magic states for the qubits in check $C_1$,
re-measure $C_1$, and continue without re-preparing and re-measuring the qubits in the other checks.
We analyze this in more detail later on specific cases.

%%%%%%%%%%%%%%%%%%%%%%%%%%%%%%%%%%%%%%%%%%%%%%%%%%%%%%%%%%%%%%%%%%%%%%%%%%%
%%%%%%%%%%%%%%%%%%%%%%%%%%%%%%%%%%%%%%%%%%%%%%%%%%%%%%%%%%%%%%%%%%%%%%%%%%%

\section{Grid Code}

The grid code is a family of outer codes where input magic states 
are laid on a square (or hypercubic) lattice points.
These have a relatively small $\nout$ compared to some other codes that we consider.

\subsection{Simple grid and pipelining}

Consider a simple hypercubic lattice, 
whose linear dimensions are $k_1, k_2, \ldots, k_D$,
on which input magic states are placed and tested.
Thus, $\nout = k_1 k_2 \cdots k_D$.
There will be checks along coordinate axes:
The first round of checks consists of 
measurements on sets of $k_1$ qubits using an $[[n_1,k_1,d_1]]$ inner code.
There are $k_2 k_3 \cdots k_D$ such measurements.
On the next round, with an $[[n_2,k_2,d_2]]$ inner code, 
there are $k_1 k_3 k_4 \cdots k_D$ measurements on sets of $k_2$ qubits.
There are $D$ rounds of checks in total.
In this subsection we calculate the order $d$ of reduction in error
without error correction,
and we are not concerned with sucess probability of the protocol.
The result is in Table~\ref{tb:gridCode}.
Note that this simple grid code as a classical code has distance $2^D$;
however, due to the sensitivity condition
the order of reduction in error for distillation purposes
is $d = 2D+1$ for $D \ge 3$
if inner codes have sufficiently high code distances.
Jones~\cite{Jones2012} used this outer code,
but it appears that he only used the fact that the order of reduction in error is at least $2D$.

For $D \ge 3$, if all the inner codes have encoding rate close to 1,
then the distillation protocol consumes $2D$ $T$-gates and $1$ $T$-state
per output magic state with the order of error reduction being $2D+1$.
Therefore, for odd $d \ge 7$ the simple grid code can be used
in place of the outer codes
defined by a biregular bipartite graph of large girth of Ref.~\cite{hhpw}.
Note that the girth of the Tanner graph of the present simple grid code 
is at most 8 regardless of $D$.

\begin{table}[b]
\begin{center}
\begin{tabular}{c|c|c}
\hline
Grid dimension $D$ & Order of error reduction $d$ & Condition \\
\hline
1 &  2 & $d_1 \ge 2$ \\
2 &  4 & $d_1 \ge 2$, $d_2 \ge 4$ \\
$D\ge 3$ & $2D+1$ & $d_j \ge 2j+1$ for all $j=1,\ldots,D$\\
\hline
\end{tabular}
\end{center}
\caption{
Order of reduction in error for the $D$-dimensional simple grid outer code.
The resulting order of error reduction in magic states is the best possible 
because when $D=1$ the outer code has code distance 2,
when $D=2$ the outer code has code distance 4,
and when $D \ge 3$ there are $D$ checks for a single qubit.
}
\label{tb:gridCode}
\end{table}

The calculation of order of reduction in error 
is inductive in the dimension of the grid.
The base case is given by a one-dimensional grid,
and the induction step by a two-dimensional grid.

We begin with the base case.
Suppose $k_1$ input magic states have independent error rate $\delta_0$.
We apply a single $H$-measurement routine
using an inner code with parameters $[[n_1,k_1,d_1]]$.
Upon successful measurement outcomes,
the output magic states will have overall error rate
\begin{align}
O(\delta_0^2 + \delta_0 \epsilon^2 + \epsilon^{d_1}).
\end{align}
The first term is due to the parity condition imposed by the outer measurement,
the second term is due to an incorrect outer code measurement,
and the last term is due to logical errors from the inner code.
It is of course important here that all error sources are independent.

Remark that 
if we decompose the output state $\mu_\mathrm{out}$ 
into the parity sectors $\Pi_\pm$ of $H^{\otimes k_1}$,
the contribution $\delta^2$ is from the even sector,
whereas $\delta \epsilon^2$ is from the odd sector.
The contribution $\epsilon^d$ due to logical errors 
may present in the both sectors.
Let us keep track of errors depending on the parity.
\begin{align}
 \delta_0'  &= \delta_0 \nonumber \\
 \delta_0'' &= 0 \nonumber \\
 \delta_1'  &= \onenorm{\Pi_- \mu_\mathrm{out} \Pi_-} = O(\delta_0 \epsilon^2 + \epsilon^{d_1})\\
 \delta_1'' &= \onenorm{\Pi_+ \mu_\mathrm{out} \Pi_+ - \mu_\mathrm{ideal} } = O(\delta_0^2 + \epsilon^{d_1})\nonumber
\end{align}
where the single prime denotes the error rate on the odd parity sector,
and the double prime on even sector.
Cross terms $\Pi_\pm \mu_\mathrm{out} \Pi_\mp$ are zero
under the stochastic error model.
Even under a general error model, they become zero after postselection
on outer measurement outcomes.

The next step (the induction step) is to consider a $k_1$-by-$k_2$ two-dimensional grid 
where each vertical column of $k_1$ qubits 
is independent and is from previous $H$-measurement routines.
($k_1$ may not be equal to $k_2$.)
As we keep track of error probabilities depending on the parity sector,
in this induction step we do not make any assumption on the magnitude of $\delta_1'$ 
compared to $\delta_1''$.
We will apply $H$-measurement routines for each row by an $[[n_2,k_2,d_2]]$ inner code.
Assume for the moment that no $H$-measurements on the rows make a logical error.
Then, the output after the row measurements has the same parity as the input state.

{\em Case 1 --- The global parity of errors is even}:
There are two subcases.
(i) 
No row measurement makes an incorrect outer code measurement.
In this case, the parity of error on each row must be even.
It is the most likely that only two rows are faulty,
which must have the same number of faulty qubits.
Hence, this has probability of order $\delta_1''^2 + \delta_1'^2$.
(ii) 
Some row measurement makes an incorrect outer code measurement.
Since this is associated with an odd parity row,
there must be at least two odd parity rows to make global parity even.
Then it is the most likely to have two errors in a single column,
or two columns of odd parity.
The probability in this case is of order $\delta_1'^2 \epsilon^4 + \delta_1'' \epsilon^4$.

{\em Case 2 --- The global parity of errors is odd}:
It is most likely that there is a single column of errors.
Then, some row measurement has to make an incorrect outer code measurement.
The probability in this case is of order $\delta_1' \epsilon^2$.

A logical error from the $[[n_2,k_2,d_2]]$ inner code
can be introduced by any of the row measurements 
regardless of whether there exists a faulty row.
Thus, a logical error increases the error probability in all cases by $O(\epsilon^{d_2})$.

In sum, we see that the error rate for the odd sector is 
$\delta_2' = O(\delta_1' \epsilon^2 + \delta_1'' \epsilon^4 + \epsilon^{d_2})$,
and that for the even sector is 
$\delta_2'' = O(\delta_1''^2 + \delta_1'^2 + \delta_1'' \epsilon^4 + \epsilon^{d_2})$.
Regarding each column as a hyperplane of a hypercube,
we obtain recursive relations of the error probabilities:
\begin{align}
 \delta_{j+1}' &= O(\delta_j' \epsilon^2 + \delta_j'' \epsilon^4 + \epsilon^{d_{j+1}} ), \nonumber\\
 \delta_{j+1}''&= O(\delta_j''^2 + \delta_j'^2 + \delta_j'' \epsilon^4 + \epsilon^{d_{j+1}}) & j \ge 0. 
\end{align}
Solving them, we arrive at Table~\ref{tb:gridCode}.

\subsection{Examples and variants}

Here and below in this section we consider specific examples.
The performance of the examples below is summarized in Table~\ref{tb:outputerror}.
For comparison, we include Table~\ref{triocost} for protocols prior to our work.
We will use an inner code with $\kin$ logical qubits. 
We take $\kin$ odd throughout.
For definiteness, 
later we consider the inner codes listed in Table~\ref{tb:innercodes}.
The numbers are from direct enumerating.
(We observed that there are several different codes with the same 
$\nin,\kin,d$ but different $\cl(d)$.)

The logical operators of the inner codes in Table~\ref{tb:innercodes},
though not shown,
are chosen such that those of weight equal to the code distance $d$
act on two or more logical qubits.
More specifically
we choose a magic basis $\ell^{(1)},\ldots, \ell^{(\kin)}$ 
of $\cS^\perp / \cS$ in regards to the self-orthogonal subspace $\cS$~\cite{hhpw}
such that
\begin{align*}
    \ell^{(a)} \cdot \ell^{(b)} &= \begin{cases} 1 & \text{if } a=b,\\ 0 & \text{otherwise,} \end{cases}\\
    |\ell^{(a)} + s| &> d \quad \forall s \in \cS.
\end{align*}
The second condition may not always be satisfied,
but for the inner codes in Table~\ref{tb:innercodes}
we randomly chose a basis of logical operators, turned it into magic basis,
and observed that there were cases where the conditions were satisfied.
This eliminates, at order $d$ and $d+1$, error patterns
where $d-1$ errors in a check becomes a logical error due to error correction
which results in the output error by one subsequent incorrect outer code measurement.

\begin{table}
    \begin{tabular}{c|c|c}
    \hline
$[[\nin,\kin,d]]$ & Stabilizer generator polynomial & $\cl(d)$ \\
    \hline
$[[31,21,3]]$ & $(x^{31}+1)/(x^5+x^2+1)$ & $155$ \\
$[[31,11,5]]$ & $(x^{31}+1)/(x^{10}+x^7+x^5+x^4+x^2+x+1)$ & $186$ \\
$[[63,45,4]]$ & $(x^{63}+1)/(x^9+x^7+x^6+x+1)$ & $1260$ \\
$[[63,39,5]]$ & $(x^{63}+1)/(x^{12}+x^9+x^7+x^5+x^3+x+1)$ & $1890$ \\
$[[63,27,7]]$ & $(x^{63}+1)/(x^{18}+x^{15}+x^{13}+x^{11}+x^9+x^5+x^4+x+1)$ & $3411$ \\ 
    \hline    
    \end{tabular}
\caption{
Inner codes. These are quantum BCH codes~\cite{qbch}.
The stabilizer group is generated by tensor products of $X$ operators and their Hadamard conjugates,
specified by a classical cyclic code generated by the given generator polynomial.
(For example, the coefficient of the first polynomial in the table 
is a binary vector in the code space of length $31$,
and any cyclic permutation is also in the code space.)
Since the code length is odd, all presented codes are normal,
and there is a choice of logical operators such that 
the transversal Hadamard is a logical Hadamard on every logical qubit.
The last column is the number of 
nontrivial $Y$-logical operators of weight equal to the code distance.
By random search,
we observed that there exists a logical operator basis
such that a logical error of weight $d$ always affects two or more logical qubits.
}
\label{tb:innercodes}
\end{table}

We take $\nout=\kin^2$ and we imagine the qubits as laid out in a two dimensional grid, 
with each qubit being labelled by a pair $(x,y)$ with $x,y \in \{0,\ldots,\kin-1\}$.

There will actually be three different families of grid codes that we consider,
depending on the checks that we take.
In the first family, the checks will consist of all verticals followed by all horizontals.
Thus, there are $\kin$ vertical checks, $C^{vert}_x$, for $x\in \{0,\ldots,\kin-1\}$, 
with $C^{vert}_x=\{(x,y)|y\in \{0,\ldots,\kin-1\}\}$
and $\kin$ horizontal checks, $C^{hor}_y$, for $y\in \{0,\ldots,\kin-1\}$, 
with $C^{hor}_y=\{(x,y)|x\in \{0,\ldots,\kin-1\}\}$.
In the second family, these vertical and horizontal checks 
are followed by diagonal checks, $C^{\searrow}_z=\{(x,y)|x+y=z \mod \kin\}$.
In the third family, we then follow those diagonal checks by additional diagonal checks, 
$C^{\nearrow}_z=\{(x,y)|x-y=z\mod \kin\}$.

All of these families of codes are $4$-cycle free.
The grid code families differ in one important way from codes that we will consider later.
The codes have distances $d=4,6,8$ for the three families respectively.
However, if one considers error patterns in which a single input qubit 
has an error and then several incorrect outer code measurements occur 
so that no outer code syndromes occur, 
these error patterns require weights $5,7,9$ respectively.
From one point of view this seems inefficient: 
By taking an even distance for the outer code, 
the error reduction is only at order $d=4$ in the first family, for example, 
even though each qubit is in two checks.
However, the weight $5,7,9$ error patterns involving a single qubit input error 
followed by several incorrect outer code measurements 
come with a large prefactor $\nout\nin^2,\nout\nin^3,\nout\nin^4$ 
for the three different families, respectively, 
while the weight $4,6,8$ error patterns which lead to an output error 
have a much smaller prefactor, and so in certain input error regimes, 
the two different types of error patterns give comparable contributions to $\eout$.

\subsection{Vertical}

The simplest grid code has
a single check on $\kin$ input $T$-states.
This outer code has distance 2.
The analysis of this simple grid code
implies that the output error probability is
$O(\einput^2 + \einput \echeck^2 + \echeck^d)$.
Thus, if the inner code for the check has distance $4$
and if we take $\einput = O(\ein^2)$, but $\echeck = \ein$,
then the output error probability becomes quartic in $\ein$.

\subsubsection{MEK $\Rightarrow [[63,45,4]]$}

For example, we can take an inner code $[[63,45,4]]$,
and can use the MEK protocol for the input $T$ state.
Then, $\einput = 9 \times 10^{-6}$ and $\echeck = 10^{-3}$.
We may employ error correction of order $m=1$ for the inner code,
but since the single check is lonely,
whenever an inner code syndrome occurs, the output will be discarded.
So, we are using the basic protocol.
The acceptance probability is $(1-\einput)^\kin (1-\ein)^{2 \nin} \approx 0.88$,
so $\nbar \approx 4.0 \times 10^1$.

The number of error patterns of the input states of weight 2
that lead to outpur error is $\binom{45}{2} = 990$,
contributing $8.0 \times 10^{-8}$ to the output error probability.
A single input error may be undetected due to incorrect outer code measurement,
the number of which is $\kin \nin = 2835$,
contributing $2.6 \times 10^{-8}$.
The contribution from the logical errors is $1.0 \times 10^{-8}$.
Thus, $\eout \approx 1.6 \times 10^{-7}$,
or $\ebar \approx 2.9 \times 10^{-9}$.
Taking the cost of the MEK protocol into account,
we have $\nTb = 5\kin + 2 \nin = 351$, or $\nTb/\nbar \approx 8.9$.

\subsection{Vertical and Horizontal}

This family of outer codes has distance $4$.
The weight $4$ error patterns which lead to no violated checks 
are all of the form of errors on the corners of a rectangle:
There are four integers, $x_1,x_2,y_1,y_2$ with $0 \leq x_1<x_2 \leq \kin-1$ 
and $0 \leq y_1 < y_2 \leq \kin-1$,
and the four errors occur on qubits $(x_1,y_1),(x_1,y_2),(x_2,y_1),(x_2,y_2)$.
Thus, there are $\binom{\kin}{2}^2$ different such error patterns.

\subsubsection{$[[31,11,5]]^{\downarrow} \Rightarrow [[31,11,5]]^{\rightarrow}$}

Taking a $[[31,11,5]]$ inner code, we have 
$\nout=121$, $\nm=22$, and $\nfin = 11$.
Taking $\ein=10^{-3}$, $\nout \ein \ll 1$,
we take $\echeck=\einput=\ein$.
We use an inner code corrected family with $m=1$
so that 
$P_{fail} = 2.0 \times 10^{-3}$ and $P_{succ} = 0.94$.
Hence, the protocol does not terminate with probability $0.85$.
% $P_{succ}/(P_{succ}+P_{fail})=0.9979\ldots$ and
% $(P_{succ}/(P_{succ}+P_{fail}))^\nm=0.955 \ldots$.
So, $\nbar \gtrsim 96$,
$\nTb \lesssim 1.5 \times 10^3$, and
$\nTb / \nbar \approx 16$.
The number of error patterns of weight $4$ leading to an output error is 
$\binom{11}{2}^2=3025$.
Thus, the error probability taking into account these terms is 
$3025 \times \ein^4 = 3.0 \times 10^{-9}$.
On the other hand, the number of error patterns of weight $5$ leading to an output error 
due to a single input error and two incorrect outer code measurements 
is equal to $\nout \times \nin^2 = 1.2 \times 10^5$ 
and the contribution of these to the output
error probability is $\nout\nin^2 \ein \echeck^4 = 1.2 \times 10^{-10}$.
All other error sources are negligible in comparison and so 
$\eout \approx 3.1 \times 10^{-9}$, 
or $\ebar \approx 3.2 \times 10^{-11}$.

\subsubsection{$[[31,21,3]]^{\downarrow} \Rightarrow [[31,11,5]]^{\rightarrow}$}

One can also consider pipelining.  
In this case, we will use two different inner codes,
a $[[31,21,3]]$ inner code on the vertical checks 
followed by a $[[31,11,5]]$ inner code on the horizontal checks.
The qubits are laid out in a {\it rectangular} two-dimensional grid.
We have now $\nout=231$.  
The values of $P_{fail} = 2.0 \times 10^{-3}$
and $P_{succ} = 0.94$ are the same as above, but $\nfin=21$, so
$\nbar \approx 161$ and $\nTb \lesssim 2.3 \times 10^3$,
or $\nTb / \nbar \approx 14$.
The number of error patterns of weight $4$ leading to an output error is
${11 \choose 2}{21\choose 2}=11550$.
Thus, the error probability taking into account these terms is 
$\approx 1.2 \times 10^{-8}$.
The number of error patterns of weight $5$ 
leading to an output error due to a single input error 
and two incorrect outer code measurements 
is equal to $231 \times 31^2$ and the contribution of these to the output
error probability is $\approx  2.2\times 10^{-10}$.
All other error sources are negligible; 
errors due to a logical error 
when measuring the vertical checks followed by an incorrect outer code measurement
on the horizontal checks 
(this has total weight $5$ since the vertical codes have distance 3)
has contribution of $2.1 \times 10^{-10}$,
and errors due to a logical error
when error correcting the vertical checks followed by one incorrect outer code measurement
on the horizontal checks does not exist due to our choice of logical operators
(this would have total weight $4$).

For the rest of the paper, 
we consider only square grid codes, rather than rectangular, 
unless otherwise mentioned.

\subsubsection{MEK $\Rightarrow [[31,11,5]]^{\downarrow} \Rightarrow [[31,11,5]]^{\rightarrow}$}

Taking the $[[31,11,5]]$ inner code for both vertical and horizontal checks,
but taking $\echeck=\ein=10^{-3}$ and $\einput=9 \times 10^{-6}$,
we find the probability that the protocol does not terminate becomes $0.95$.
The error patterns with a single input error 
and two incorrect outer code measurements contribute $\approx 1.0 \times 10^{-12}$ 
to the output error probability,
while error patterns with a logical error contribute $\approx 3.3 \times 10^{-11}$ 
to the output error probability, with all other contributions negligible 
% so that $\eout \approx 3.1 \times 10^{-14}$.
so that $\ebar \approx 2.8 \times 10^{-13}$.
We have $\nbar \gtrsim 1.1 \times 10^2$.
% $\nTb \lesssim 1.5 \times 10^3$.
In the checks we consume $1.4 \times 10^3$ $T$-gates,
and the MEK protocol to have $\einput=9\times 10^{-6}$
consumes $5$ $T$-gates per $T$-state input to the vertical checks.
Overall, with the MEK protocol included, $\nTb / \nbar \approx 19$.

\subsubsection{MEK $\Rightarrow[[31,11,5]]^{\downarrow}\Rightarrow [[31,11,5]]^{\rightarrow}_{MEK}$}

We now further modify the protocol:
For the vertical checks, we will use an error of $10^{-3}$,
while for the input magic states and the horizontal checks we will use $9 \times 10^{-6}$.
The idea is that for the vertical checks, which are done first,
one may tolerate a higher $T$-gate error, 
since any logical error that is produced will likely be caught by the horizontal checks.
The dominant error pattern is a single input error, 
followed by two incorrect outer code measurements; 
there are $\nout \nin^2$ such error patterns 
contributing an error $\approx 8.5 \times 10^{-17}$,
with the next most significant error pattern being four input errors 
contributing $\approx 2.0 \times 10^{-17}$ 
so one finds that $\eout \approx 1.0 \times 10^{-16}$.
The repeat probability for horizontal checks
is negligibly small ($5.6\times 10^{-4}$),
whereas that of vertical checks is $6.2 \times 10^{-2}$.
Hence we find that $\nbar \approx 1.2 \times 10^2$ and
$\ebar \approx 8.9 \times 10^{-19}$.
% For the vertical checks, we have  $P_{fail}=31*10^{-6}+(.022)^2/2+O(\echeck^3)\approx 0.0003\ldots $ and $P_{succ}=0.969 \ldots$ with
% $P_{succ}/(P_{succ}+P_{fail})^\nm=0.9938\ldots$.
% Thus, $\nbar\approx 121*0.9938\approx 116$, 
% since the repeat probability for the horizontal checks is negligibly small (i.e., output states are very likely to be high quality).
Assuming success,
the number of input magic states with error $10^{-3}$ is $62 \cdot 11 = 682$,
while the number with error $9 \times 10^{-6}$ is $682 + 121 = 803$.
Taking the cost of MEK protocol (and the repetition and failure probability) into account,
$4.7 \times 10^3$ magic states with error $10^{-3}$ are consumed on average.
Per output, this is $\approx 40$ input $T$'s.
% Thus, it requires
% $5*803+121=4356$ input magic states with error $10^{-3}$ to produce the output, or roughly $36$ input states per output state.
% WHY EBAR IS REALLY SA

\subsubsection{
$[[63,39,5]]^{\downarrow} \Rightarrow [[63,39,5]]^{\rightarrow}$ with $\ein = 9\times 10^{-6}$
}

Now we consider a $[[63,39,5]]$ inner code.
We have $\nout=1521$ and $\nm=78$.
We take $\einput=\echeck = 9\times 10^{-6}$.
We use an inner code corrected protocol with $m=1$,
though a basic protocol without error correction performs similarly.
% so that $P_{fail}\approx 6\times 10^{-7} $.  We have $P_{succ}=0.9988 \ldots$ with
% $(P_{succ}/(P_{succ}+P_{fail}))^\nm=0.9999 \ldots$.
% Hence,
% $\nbar\approx 1521$
We find that $\nbar \approx 1.5 \times 10^3$ 
and $\nTb \approx \nout + 2\nin \nm /(1-P_{repeat}) \approx 1.1 \times 10^4$.
Thus, $\nTb/\nbar\approx 7.4$.
There is a factor of $5$ overhead in $T$ count 
to distill the input gates with error $9\times 10^{-6}$ from gates with error $10^{-3}$,
so it requires $\approx 37$ input magic states with error $10^{-3}$.
The number of error patterns of weight $4$ leading to an output error is 
$\binom{39}{2}^2=5.5 \times 10^5$.
Thus, the error probability taking into account these terms 
is $\binom{39}{2} \ein^4 \approx 3.6\times 10^{-15}$, 
so $\ebar \approx 2.4 \times 10^{-18}$.
All other error sources are negligible in comparison.

\subsubsection{
MEK    $\Rightarrow [[63,39,5]]^{\downarrow}\Rightarrow [[63,39,5]]^{\rightarrow}_{MEK}$
}
As before with the $[[31,11,5]]$ inner code,
we can modify the protocol where we use raw $T$ gates 
in the first round of checks (vertical) $\echeck = 10^{-3}$,
but in the input $T$-states and the second round of checks (horizontal)
we use distilled $T$-gates at error rate $9 \times 10^{-6}$ from the MEK protocol.
This reduces the number of raw $T$ states required.
We perform all vertical checks in parallel, 
and only if they succeed do we perform the horizontal checks. 
For the vertical checks, we have $P_{fail}=8.0\times 10^{-3}$
and $P_{succ}=0.88$ with $P_{succ}/(P_{succ}+P_{fail})=0.991$.
There are $39$ vertical checks, with $0.991^{39} =0.70$.
Thus, $\nbar \approx 1521*0.70 \approx 1.1 \times 10^3$,
where we have neglected the probability of error correction in the horizontal checks
since it is negligibly small.
The output error contribution from quadruples of input $T$-state error
is $3.6 \times 10^{-15}$,
which is now comparable with the contribution of $4.4 \times 10^{-15}$
from error patterns of weight 5 where a single input $T$-state error (weight 1) 
is combined with two incorrect outer code measurements (weight 4).
Logical errors are negligible in comparison.
Hence, $\ebar \approx 7.6 \times 10^{-18}$.
Taking the cost of the MEK protocol and the repeat and failure probabilities,
we find
$\nTb \approx 2.9 \times 10^4$ and $\nbar \approx 1.1 \times 10^3$,
or $\nTb / \nbar \approx 28$.

\subsection{Vertical, Horizontal, and One Diagonal}

The second family of outer codes has distance $6$.  
The weight $6$ error patterns which lead to no violated checks are all of the following form.
There are three integers $x,y,l$ and the six errors occur on distinct qubits
$(x-l,y-l),(x,y-l),(x-l,y),(x+l,y),(x,y+l),(x+l,y+l)$.
This error pattern corresponds to taking two squares which share a corner, 
and including all qubits at the corners of the squares except for the shared corner.
There are in total $\kin^2 (\kin-1)/2$ such patterns.

\subsubsection{
MEK 
$\Rightarrow [[63,27,7]]^\downarrow \Rightarrow [[63,27,7]]^\rightarrow \Rightarrow [[63,27,7]]^\searrow$
}

Taking a $[[63,27,7]]$ inner code, we have $\nout=729$ and $\nm=81$.
If we take $\echeck=\einput=\ein=10^{-3}$,
we have $(1-\ein)^\nout=0.482$.
Since this number starts to become small,
it is worth instead taking $\echeck=10^{-3}$ 
and $\einput = 9 \times 10^{-6}$ so $(1-\einput)^\nout= 0.994$.
While this increases the number of physical $T$-gates required 
if one gets the given $\einput$ by a quadratic distillation protocol,
it is compensated by the increased success probability.

For an inner code corrected family with $m=1$,
we have 
$P_{fail} = \nin \echeck^2 + 2\nin^2 \echeck^2 +O(\echeck^3) \approx 0.008$, 
$P_{succ} = (1-\echeck)^{2\nin} \approx 0.882$,
$P_{repeat} = 2\nin \echeck \approx 0.126$
so
$P_{succ}/(P_{succ}+P_{fail}) = 0.991$.

First consider the measurement of vertical checks.
We repeat each vertical measurement until it succeeds,
only reinitializing the qubits in that check.
With probability
$p_1 = P_{succ}/(P_{succ}+P_{fail}) = 0.991$, 
we succeed in measuring this check without an outer code syndrome or inner code syndrome 
that we cannot correct.
With probability $1-p_1$,
we must reinitialize and remeasure the qubits on the check.
For each vertical check, it requires $27$ input magic states and
on average $2 \nin / (1-P_{repeat}) \approx 144$ $T$-gates.
Repeating each check until we succeed, it requires
$27 / p_1 \approx 27.2$ input magic states and $144/p_1 \approx 145$ $T$-gates per vertical line.

After measuring the vertical checks, 
we then measure horizontal and diagonal checks.
The method that uses the fewest $T$-gates is to measure these checks sequentially,
terminating if any fail.
However, since this requires a fairly large time overhead compared to a parallel method,
we instead measure all horizontal checks in parallel,
and then if all succeed, we measure all diagonal checks in parallel;
if any horizontal checks fail, we do not measure the diagonal checks.

The probability of success on all horizontal and diagonal checks 
is $\approx p_1^{2 \nm /3} =0.614$.
Thus, the average number of output magic states is 
$\nbar \approx (729-\nfin P_{repeat} \kin) p_1^{2\nm/3} \approx 391$
The average number of used input magic states at error rate $9 \times 10^{-6}$ is $\approx 736$.
The average number of $T$-gates used is
$\approx 145 \times 27$ for the vertical checks 
and $\approx 144 \times 27$ for the horizontal or diagonal checks.
The probability that all horizontal checks succeed is $p_1^{\nm/3} \approx 0.784$,
and, with this probability, we then measure all diagonal checks.
Thus, the total number of $T$-gates used is 
$ 145\cdot 27 + 144 \cdot 27 + 0.784 \times 144 \cdot 27 = 1.1 \times 10^4$,
in addition to the $\approx 5 \cdot 736$ input magic states at $10^{-3}$ error rate.
Hence, the total $T$-gate/state count is $1.5 \times 10^4$,
or $\nTb/\nbar \approx 37$.

The number of error patterns of weight $6$ leading to an output error is 
$\kin^2 (\kin-1) /2 = 9477$.
These lead to a negligible contribution to the output error due to the much smaller $\einput$.
On the other hand, the number of error patterns of weight $7$ 
leading to an output error due to a single input error 
and three incorrect outer code measurements is
$\nout \times \nin^3 =1.8 \times 10^8$
and the contribution of these to the output
error probability is $\approx 1.6 \times 10^{-15}$.
The number of error patterns of weight $7$ due to an inner code logical error 
leading to an output error is $2^6 \cdot \nfin \cl(d) = 5.9 \times 10^6$
and these lead to a contribution to the output error probability of 
$5.9 \times 10^{-15}$.
All other error sources are negligible in comparison 
and so $\eout \lesssim 7.5 \times 10^{-15}$,
or $\ebar \approx 1.9 \times 10^{-17}$.

\subsubsection{
$[[63,27,7]]^\downarrow \Rightarrow [[63,27,7]]^\rightarrow \Rightarrow [[63,27,7]]^\searrow$
with $\ein = 9 \times 10^{-6}$.
}
As before, $\nout = 729$, $\nm = 81$, and $\nfin = 27$,
and we use $m=1$ error correction for inner codes.
At error rate $\echeck = \einput = \ein = 9 \times 10^{-6}$,
we have $P_{succ} = 0.999$, $P_{fail} = 6.5 \times 10^{-7}$, 
and $P_{repeat} = 1.1 \times 10^{-3}$,
so the acceptance probability is $0.993$,
even without considering partial restart.
(The acceptance probability using the basic protocol without inner code correction
is $(1-\ein)^{\nout + 2\nin \nm} = 0.906$.)
Using the combinatorial factors we computed above for error patterns that lead to output errors,
we see that the contribution from errors of weight 6 on the input $T$ states is 
$5.0 \times 10^{-27}$,
that of weight 7 from single input errors combined with three 
incorrect outer code measurements is
$8.7 \times 10^{-28}$,
and that from the logical errors in lonely checks is
$2.8 \times 10^{-29}$.
so $\eout \approx 5.9 \times 10^{-27}$.

Discarding lower quality qubits, we have $\nbar \approx 7.2 \times 10^2$,
and the number of $T$-states/gates consumed is $1.1 \times 10^4$ at error rate $9\times 10^{-6}$.
Hence, $\nTb/\nbar \approx 15$, 
or $\approx 75$ including the $T$-count of initial distillation 
by the MEK protocol.
We have $\ebar \approx 8.2 \times 10^{-30}$.

\subsection{Vertical, Horizontal, and Both Diagonals}
The grid code with a vertical, horizontal, and both diagonal checks gives us an outer code with distance $8$.  
However, in order to suppress other errors at a comparable order, we need an inner code with distance $9$.  One possible candidate code is a $[[73,19,9]]$ inner code.  However, this consumes a large number of $T$-gates and the success probability becomes quite small at input error rates around $10^{-3}$ unless we use an unacceptable amount of error correction.  It may be worth considering this code for other error rates but we do not discuss it further here.

\section{Graph Outer Codes}

\subsection{$d=5$}

The sensitivity requires that each qubit should be in at least two checks.
Hence, we consider graphs where a qubit corresponds to an edge, 
and a check corresponds to a vertex of degree $\kin$,
with a qubit $q$ in some check $C$ 
if the corresponding edge is attached to the corresponding vertex.  
Thus each qubit is in two checks.
The outer code's distance is the girth of this graph.
The Petersen graph code~\cite{hhpw} is an example of this for distance 5.

The literature gives several examples of small graphs with fixed degree and girth $5$.
For degree $7$, the smallest graph with girth $5$ 
is known to be the Hoffman-Singleton graph~\cite{hoffman1960moore}.
This graph has $50$ vertices and $50*7/2=175$ edges.
For degree $9$, the smallest graph with girth $5$ 
that we could find in the literature~\cite{jorgensen2005girth,adjacency} 
has $96$ vertices and $96*9/2=432$ edges.
For degree $11$, the smallest we could find~\cite{jorgensen2005girth,adjacency} 
has $156$ vertices and $156*11/2=858$ edges.

The Hoffman-Singleton graph has $1260$ $5$-cycles,
while the degree $9$ graph has $8960$ $5$-cycles,
and the degree $11$ graph has $24336$ $5$-cycles.

These graph codes can be used with inner codes such that $\kin - \text{degree}$
is a nonnegative even number.
These codes differ from the grid code above,
in that the outer code distance $5$ is the same as the order 
at which errors occur due to a single input error 
followed by incorrect outer code measurements of every check involving that input magic state 
(i.e., also $5$).
Thus, these codes allow fifth order reduction for the given $\kin$ 
with a smaller $\nT$ than any other code that we know.
However, we do not analyze these codes with specific numbers 
because it appears that in many practical regimes the grid codes 
will give better performance 
(at sufficiently small input error these graph codes may become superior).

\subsection{$d=7$}

For $d=7$, 
we consider a family of outer codes which slightly generalize those in Ref.~\onlinecite{hhpw}.
In Lemma~9 of that reference, 
it was shown that for any $\kin$, 
for any odd distance $d\geq 5$, 
for sufficiently large $\nout$, 
one can obtain an outer code with the distance and sensitivity properties 
so that every qubit is in exactly $(d-1)/2$ checks.
For the case $d=7$, this requires that each qubit be in $3$ checks.
Here, we consider how to do this with as small $\nout$ as possible;
in some cases, we do this by slightly increasing the number of checks
so that some small fraction of qubits are in more than $(d-1)/2$ checks.

We choose the Tanner graph to be such that
all checks to have degree $\kin$ and all qubits to have degree $3$.
We now show that any such Tanner graph with girth $6$ or more 
(i.e., the code is $4$-cycle free)
and which defines a code with distance $7$ or more 
will give an outer code that obeys the distance and sensitivity bounds.
First, any single input error will violate $3$ checks,
since every bit is in three checks.
Any pair of input errors on qubits $q_1,q_2$ must violate at least $4$ checks
(since each qubit is in $3$ checks, and the code is $4$-cycle free, 
there is at most one check containing both $q_1,q_2$).
Any three input errors on qubits $q_1,q_2,q_3$ 
must also violate at least $3$ checks 
(there is at most one check containing $q_1,q_2$ 
and at most one check containing $q_2,q_3$ 
and at most one check containing $q_1,q_3$).
Any four input errors on qubits $q_1,q_2,q_3,q_4$ 
must violate at least $2$ checks 
(the number of violated checks must be even since there are an even number of input errors,
and by the distance assumption, there is no pattern on four qubits that violates no checks).
By the distance assumption, any five or six input errors must violate at least one check.

We performed a numerical search for graphs with the needed girth which defined a code 
with the needed distance as follows: 
We chose an integer $\alpha$ and 
searched an outer code with $\alpha \kin$ qubits and $3\alpha$ checks.
The search was an iterative randomized procedure.
We initialized the graph by taking $\alpha$ copies of the complete bipartite graph 
on $\kin$ qubits and $3$ checks.
This initial graph has girth 4.
We then performed an iterative random search 
to find a graph with girth $6$ or larger;
this search proceeded by first finding a $4$-cycle,
then choosing an edge $(q,C)$ between a qubit $q$ and an edge $C$ in that $4$-cycle,
then choosing another random edge $(q',C')$ 
and replacing the pair $(q,C)$ and $(q',C')$ with $(q,C')$ and $(q',C)$.
This procedure was repeated until the graph had girth $6$ or larger.
Then, an additional random update was performed;
this update also replaced pairs of edges $(q,C)$ and $(q',C')$ with $(q,C')$ and $(q',C)$;
in this case, the pairs were chosen randomly subject to the constraint 
that no $4$-cycle is created.
After a large number of such steps,
we tested whether the resulting code had distance $7$;
this test was done by searching for an error pattern of weight $6$ or less 
that does not violate an outer code check;
some tricks were done to speed this search 
(for example, if a qubit $q$ has an error, 
and if $q$ is in checks $C_1,C_2,C_3$ 
then there must be qubits $q_1\in C_1,q_2\in C_2,q_3\in C_3$ with $q_1,q_2,q_3\neq q$ 
such that $q_1,q_2,q_3$ all have errors).

For $\kin=5,7,9,11,13$, for $\alpha=\kin+1$,
we were able to find graphs with girth $6$ by random search.
Note that there exist graphs with girth $6$ with $m=\kin$ 
(the grid code with horizontal, vertical, and one diagonal is an example of such),
but we did not find them. 
However, we did not find graphs with both girth $6$ and distance $7$ until a larger $\alpha$.
These graphs give concrete examples of outer codes 
which obey the distance and sensitivity bounds.

We also found outer codes which obey the distance and sensitivity bounds 
with $\nout = \alpha \kin$ qubits for smaller values of $m$ by taking more checks.
We did this as follows:
We first found graphs of girth $6$ or more as described above 
and then did a large number of random updates of these graphs keeping girth $\geq 6$.
Then, if the resulting code had distance $5$ or $6$,
we tried to find whether one could add a small number of checks to that code 
to obtain a code with distance $7$.
The resulting code then obeys the distance and sensitivity bounds.

The results of these searches are shown in Table~\ref{minm}.
Thus, these codes allow seventh order reduction in error 
with a smaller $\nT$ for the given $\kin$ than any other code that we know.
However, as in the case of graph codes, we do not analyze them further.
\begin{table}[hb]
\begin{tabular}{c|c|c}
    \hline
$\kin$ & $\alpha$ & $\alpha$ with added checks\\
\hline
5 & 7 \\
7 & 13 &10 \\
9 & 19 & 14 \\
11 & 33 & 20\\
13 & 45 & 29\\
\hline
\end{tabular}
\caption{
Outer codes $M$ such that $2|Me|+|e| \ge 7$ found in randomized search.    
For given degree equal to $\kin$, 
the second column labelled $\alpha$ shows the minimum $\alpha$ 
at which we found a constant degree Tanner graph 
giving a code obeying the distance and sensitivity bounds.
(These have the optimal number of checks per output at 7$^{th}$ order of reduction in error.)
The third column $\alpha$ with added checks 
shows the minimum $\alpha$ giving a code obeying the weight and sensitivity bounds 
where we add one or two checks to a constant degree Tanner graph.
In any case, $\nout = \alpha \kin$.
} 
\label{minm}
\end{table}

\begin{table}[p]
\begin{tabular}{c|c|c|c|c}
\hline
Protocol & $\nout$ & $\nbar$ & $\ebar$ & $\nTb/\nbar$\\
\hline

$MEK \Rightarrow [[63,45,4]]$ &
$45$ & $4.0 \times 10^1$ & $2.9 \times 10^{-9}$ & $8.9$ \\

$[[31,11,5]]^\downarrow \Rightarrow [[31,11,5]]^\rightarrow$ &
$121$ & $9.6 \times 10^1 $ & $3.2 \times 10^{-11}$ & $16$ \\

$[[31,21,3]]^\downarrow \Rightarrow [[31,11,5]]^\rightarrow$ &
$231$ & $1.6 \times 10^2$ & $7.5 \times 10^{-11}$ & $14$ \\

$MEK \Rightarrow [[31,11,5]]^\downarrow \Rightarrow [[31,11,5]]^\rightarrow$ &
$121$ & $1.1 \times 10^2$ & $2.8\times 10^{-13}$ &  $19$ \\

$MEK \Rightarrow [[31,11,5]]^\downarrow \Rightarrow [[31,11,5]]^\rightarrow_{MEK}$ &
$121$ & $1.2 \times 10^2$ & $8.9 \times 10^{-19}$ &  $40$ \\

$MEK \times \left( [[63,39,5]]^\downarrow \Rightarrow [[63,39,5]]^\rightarrow \right)$ &
$1521$ & $1.5 \times 10^3$ & $2.4\times 10^{-18}$ & $37$ \\

$MEK \Rightarrow [[63,39,5]]^\downarrow \Rightarrow [[63,39,5]]^\rightarrow_{MEK}$ &
$1521$ & $1.1 \times 10^3$ & $7.6 \times 10^{-18}$ & $28$ \\

$MEK \Rightarrow [[63,27,7]]^\downarrow_{\circlearrowleft} 
\Rightarrow [[63,27,7]]^\rightarrow \Rightarrow [[63,27,7]]^\searrow$ &
$729$ & $3.9 \times 10^2$ & $1.9 \times 10^{-17}$  & $37$ \\

$MEK \times \left( [[63,27,7]]^\downarrow 
\Rightarrow [[63,27,7]]^\rightarrow \Rightarrow [[63,27,7]]^\searrow \right)$ &
$729$ & $7.2 \times 10^2$ & $8.2 \times 10^{-30}$  & $75$ \\

\hline
\end{tabular}
\caption{
Output Error of protocols using a grid outer code,
where qubits are placed on a two-dimensional grid,
assuming $\ein=10^{-3}$ and perfect Clifford operations.
Order $m=1$ error correction for inner codes is used.
``$MEK \Rightarrow$'' means that the input magic states 
are from the MEK (10-to-2) protocol~\cite{MEK},
whereas ``$MEK \times$'' means that all $T$-states/gates are from the MEK protocol.
The subscript $MEK$ means that the $T$-gates in those checks are from the MEK protocol.
The column $\nTb/\nbar$ includes $T$ cost of the MEK protocol.
The subscript $\circlearrowleft$ means that a column is started over until success.
The superscript arrows indicate the direction of the checks.
$\nout$ is generally larger than $\nbar$,
because failure probability is nonzero and lower quality outputs are discarded.
The space requirement is $c ~\nout$ (noiseless) qubits where $ 1 \le c < 4$.
}
\label{tb:outputerror}
\end{table}

\begin{table}[pb]
\begin{tabular}{c|c|c|c}
\hline
Protocol & $\nout$ & $\eout_{marginal}$ & $\nTb/\nbar$ (Space cost) \\
\hline
6-22-54 & 7128 & $1.0 \times 10^{-13}$ & 47 ($3.3 \times 10^5$)  \\
$\star$ 15-5    & 2 &$1.1 \times 10^{-14}$ & 76 ($1.5 \times 10^2$) \\
5-34-46 & 3128 &$9.8 \times 10^{-15}$ & 52 ($1.6 \times 10^5$) \\
% 5-14-26 & 728 &$9.8 \times 10^{-16}$ & 60 ($4.3 \times 10^4$) \\
5-5-54 & 216 &$8.7 \times 10^{-17}$ & 80 ($1.7 \times 10^4$) \\
$\star$ 5-5-5 & 8 &$4.8 \times 10^{-18}$ & 126 ($1.0 \times 10^3$)\\
22-46-54-54 & 2950922($\approx 3 \times 10^6$) &$8.1 \times 10^{-19}$ & 115 ($3.1 \times 10^8$) \\
$\star$ 15-15 & 1 &$1.5 \times 10^{-21}$ & 228 ($2.3 \times 10^2$) \\
\hline
\end{tabular}
\caption{
Costs of protocols using concatenations of triorthogonal $(3k+8)$-to-$k$ protocol (BH)~\cite{BravyiHaah2012Magic}
denoted by an even number $k$,
MEK~\cite{MEK} denoted by ``5'', and the 15-qubit code (BK)~\cite{bk} denoted by ``15'', 
starting with $\ein = 10^{-3}$ input $T$-states.
$T$-counts are rounded from the first decimal place.
The space cost is the number of qubits required to run the protocol,
ensuring the independence of the magic states to the next round of protocol;
for example, in ``6-22-54'', $(3 \cdot 6 + 8)(3\cdot 22 + 8)(3 \cdot 54 + 8) = 3.3 \times 10^5$
initial $T$-states are needed.
We assumed MEK protocol's 10 $T$ states are occupying 10 qubits.
The protocols are chosen according to the lowest $T$-count 
given a target marginal output error rate using BH with $k\le 54$, MEK, and BK.
Those with $\star$ are similarly chosen, but using only MEK and BK.
All error rates are computed by marginal error probability at each stage,
so the actual error rate might be lower; see~\cite{BravyiHaah2012Magic} for more detail.
}
\label{triocost}
\end{table}

%\pagebreak
\bibliography{enum-ref}
\end{document}